%% file: main.tex
\pdfoutput=1
\documentclass[12pt,a4paper]{article}

\usepackage{ifthen} 
\newboolean{pdflatex}
\setboolean{pdflatex}{true} 

\usepackage{placeins}
\usepackage{subfig}
\newboolean{articletitles}
\setboolean{articletitles}{true} 

\newboolean{uprightparticles}
\setboolean{uprightparticles}{false} 

\def\paperauthors{LHCb collaboration} 
\def\paperasciititle{Heavy ion physics with LHCb Upgrade II} 
\def\papertitle{Heavy ion physics with\\LHCb Upgrade~II} 
\def\paperkeywords{ {LHCb}} 
\def\papercopyright{\the\year\ CERN for the benefit of the LHCb collaboration} 
\def\paperlicence{CC BY 4.0 licence}
\def\paperlicenceurl{https://creativecommons.org/licenses/by/4.0/}

\input{preamble}
\usepackage{comment}
\usepackage{tikz}
\usetikzlibrary{shapes}

\begin{document}

\newcommand{\pp}{\proton\proton}
\newcommand{\pPb}{\proton{}Pb}
\newcommand{\PbPb}{\text{PbPb}\xspace}

\renewcommand{\thefootnote}{\fnsymbol{footnote}}
\setcounter{footnote}{1}
\input{title-LHCb}
\renewcommand{\thefootnote}{\arabic{footnote}}
\setcounter{footnote}{0}

\cleardoublepage


\pagestyle{plain} 
\setcounter{page}{1}
\pagenumbering{arabic}


\section{Introduction}

The successful operation and rich physics harvest of \lhcb during Run~1 and Run~2 of the \lhc has vindicated the concept and design of a dedicated heavy-flavour physics experiment at a hadron collider. 
This has inspired efforts to upgrade the experiment's capabilities, to operate at higher instantaneous luminosities and fully exploit the unique physics opportunities offered by the unprecedented rate of heavy quarks that \lhc collisions produce at full luminosity.
While the initial \lhcb experiment aimed at an instantaneous $pp$ luminosity of $2\times 10^{32}\cm^{-2}\sec^{-1}$, the Upgrade~I of the detector was designed and has recently been proven to work at $2\times 10^{33}\cm^{-2}\sec^{-1}$.
The \lhcb collaboration is now looking at another order of magnitude increase, aiming at luminosities in excess of $10^{34}\cm^{-2}\sec^{-1}$, with its Upgrade~II project. 
Detailed plans for \lhcb Upgrade~II have been set out in a Framework Technical Design Report~\cite{LHCb-TDR-023}, supplemented by a Scoping Document~\cite{LHCb-TDR-026}.
The physics programme is described at length in a separate document~\cite{LHCb-PII-Physics}.

The unique features of the \lhcb detector, specifically its forward geometry (covering the pseudorapidity range $2 < \eta < 5$ with acceptance down to low \pt values) and its capability to identify cleanly strange, beauty and charm hadrons in a range of final states, also provide excellent potential to study interesting phenomena in heavy ion collisions.  
A rich programme has already been developed with data from Run~1 and Run~2, but in \PbPb collisions the centrality that can be probed at \lhcb is, until now, limited by the granularity of the detectors.  
The \lhcb Upgrade~II tracking system will allow to overcome this limitation.
It will consist of a vertex locator (VELO) surrounding the interaction region followed by tracking stations both upstream and downstream of the \lhcb dipole magnet.
Silicon pixels will be used for the VELO and the Upstream Pixel (UP) trackers, as well as for the central region of the downstream system (called the Mighty Tracker, MT).  
The outer region of the MT, where the occupancy is lower, will be instrumented with scintillating fibre technology.
A similar scintillating fibre technology is proposed for tracking stations on the magnet side walls (Magnet Stations, MS), increasing the acceptance for low \pt\ tracks.  
Together, this system will allow reconstruction of $pp$ collisions at a peak pile-up of 30--40 or even higher.
Extrapolating from the limitation in centrality achieved by the original and current LHCb detectors, and verified by simulation studies, the LHCb Upgrade~II detector design supports the reconstruction of the most central \PbPb collisions.

This document provides a summary of the heavy ion physics potential at \lhcb Upgrade~II, drawing on content presented at a recent dedicated workshop on this topic~\cite{LHCb-PUB-2024-007}.
Recently, heavy ion running for about one month per year after LS4 has been established in the High-Luminosity LHC (HL-LHC) schedule, and it is assumed in what follows that the programme will be similar to that pursued to date: a significant amount of running time dedicated to \PbPb collisions supplemented with additional runs in other configurations, including asymmetric systems such as $p$Pb and, potentially, small collision systems such as OO. 
Higher instantaneous luminosity than achieved to date will allow much larger data samples to be accumulated (an integrated luminosity of $10\invnb$ of PbPb collisions is assumed for projections); unlike in $pp$ collisions this will be achieved with a collision pile-up significantly lower than one.
The \lhcb Upgrade~II detector will have the capability to allow important measurements to be made from a range of different heavy ion beam species, which in some cases could extend the physics programme even beyond what is described in this document.

In addition to ion-ion, proton-ion and ion-proton beam collisions, \lhcb has uniquely demonstrated the capability to operate in fixed target mode, achieved by injecting gases into the interaction region~\cite{Bursche:2018orf,LHCB-TDR-020,LHCb-DP-2024-002}.
A further upgrade~\cite{LHCspin2024} of the injection system could allow studies of interactions between beam particles and polarised hydrogen or deuterium gases, enabling a rich programme of spin physics that expands the physics reach of the \lhc complex.
This potential extension to the \lhcb Upgrade~II physics programme is not described here, but would provide important complementary information on several of the topics discussed.

\section{Partonic structure of nuclei/nucleons}

\begin{figure}[tb]
\centering
\includegraphics[width = 0.95\textwidth]{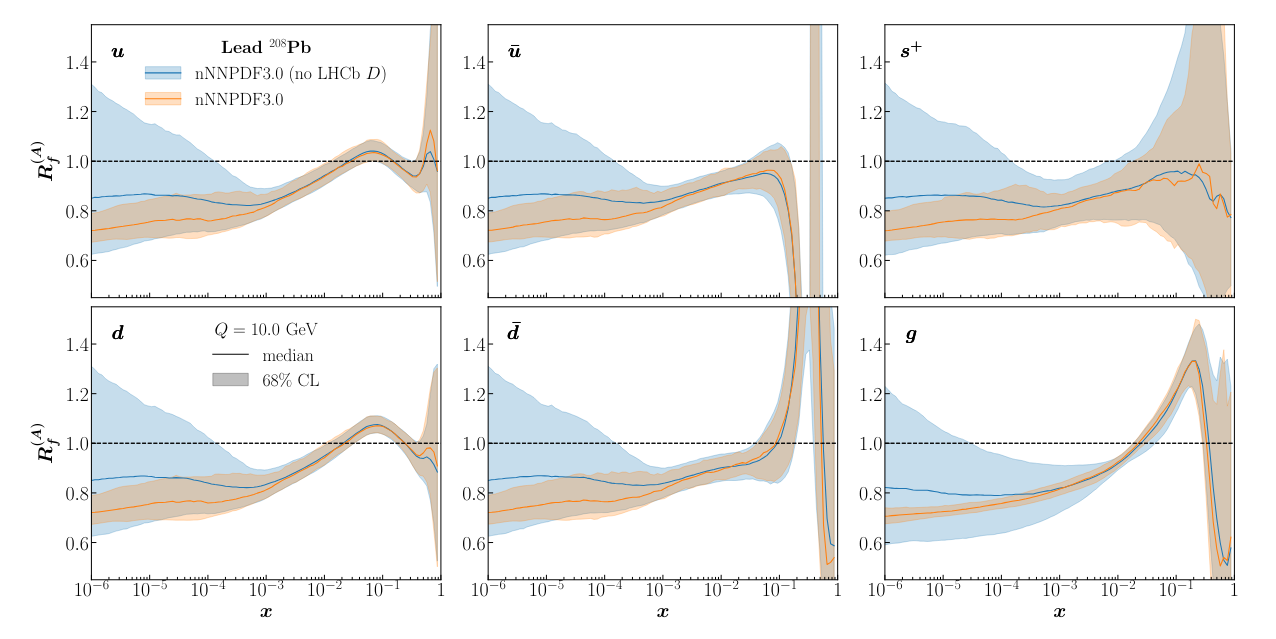}
\caption{Impact of the \lhcb\ measurement of \Dz production in \textit{p}Pb collisions~\cite{LHCb-PAPER-2017-015} on the nuclear PDFs, as a function of Bjorken $x$, according to the nNNPDF3.0 model~\cite{AbdulKhalek:2022fyi}.}
\label{fig:nPDF}
\end{figure}

The unique kinematic coverage of the fully instrumented \lhcb spectrometer provides high-precision measurements of identified particle production in ion collisions at far forward rapidity. Such measurements constrain models of the partonic structure of nuclei at the lowest values of Bjorken~$x$ that are experimentally accessible. Currently, \lhcb measurements of \Dz meson production at forward rapidity in $p$Pb collisions provide the primary constraint on the gluon structure functions for $x<10^{-4}$.  In particular, Fig.~\ref{fig:nPDF} shows that uncertainties on nuclear parton distribution functions (nPDFs) are dramatically reduced and constrained down to $x=10^{-6}$ by the inclusion of \lhcb data.  Despite this progress, signatures of the gluon saturation regime remain elusive. The \lhcb collaboration will continue to pursue a vigorous search for the onset of saturation, including measurements of forward direct photons, photoproduced charm, and identified hadrons, taking advantage of the improved performance offered by the Upgrade~II detector.  Measurements of relatively high $Q^{2}$ processes, such as $Z^{0}$ boson production, will also be pursued to constrain nPDFs.

\begin{figure}
\centering
\includegraphics[width = 0.75\textwidth]{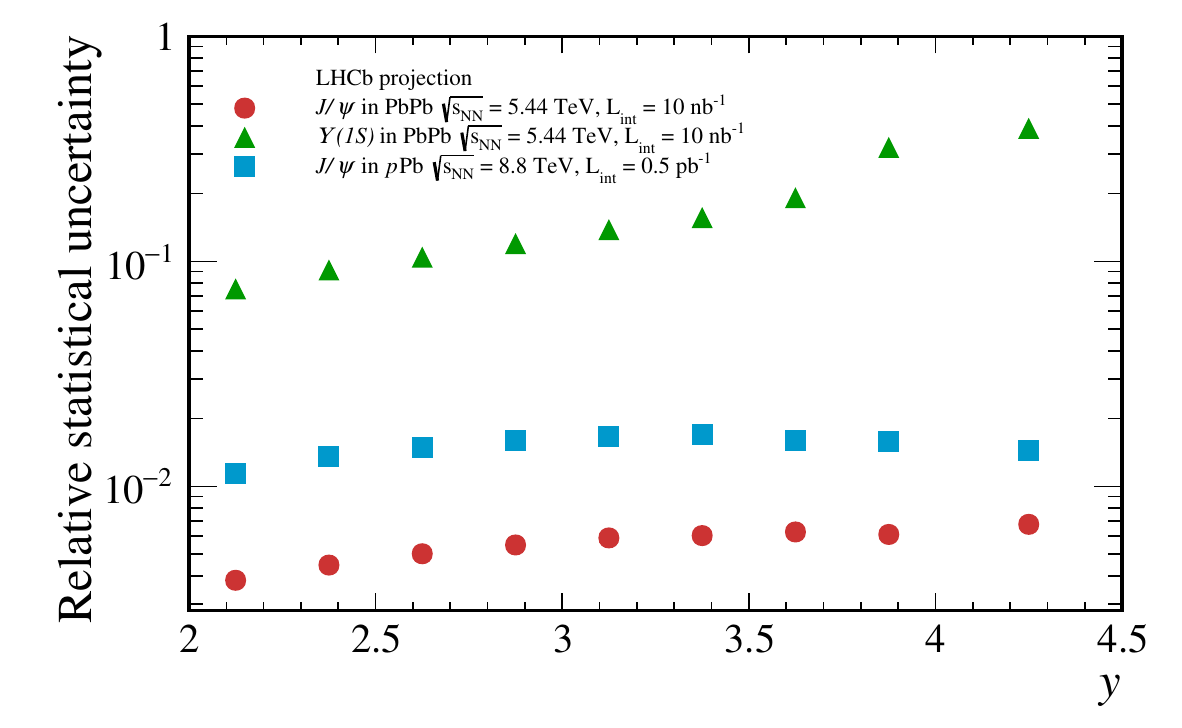}
\caption{Projected uncertainties for measurements of exclusive \jpsi and  \OneS production in \PbPb collisions at $\sqsnn=5.44\tev$ and of exclusive \jpsi production in \textit{p}Pb collisions at $\sqsnn=8.8\tev$, as a function of rapidity ($y$).}
\label{fig:UPC}
\end{figure}

Complementary information about the nucleon and nuclear structure is given by the study of Ultra-Peripheral Collisions (UPC), where interactions are mediated by the exchange of photons or QCD colourless objects, such as pomerons or odderons. 
When a hard scale is involved in the process under study, \eg\ through the production of a particle with a large enough mass, the measurements are sensitive to various aspects of the hadron structure: exclusive processes provide access to generalised parton distributions~\cite{Diehl:2003ny}, dissociative diffractive production provides sensitivity to local fluctuations in the gluon distribution~\cite{Mantysaari:2017dwh}, and inclusive dijet production probes collinear parton distribution functions. 
These processes potentially also offer sensitivity to saturation effects~\cite{Armesto:2014sma,Fucilla:2023mkl}. 
Data collection at large luminosity with the \lhcb Upgrade~II detector would allow the measurement of exclusive processes at \lhcb to be multi-differential in rapidity and momentum transfer to the final-state beam hadron, setting strong constraints on the generalised parton distributions. 
Moreover, the study of different final states, such as $\phi$, \jpsi, and \Upsilonres mesons, provides measurements at different scales and allows QCD evolution effects to be probed. 
Figure~\ref{fig:UPC} shows projections for the statistical uncertainty, scaled from simulation and previous results~\cite{LHCb-PAPER-2024-012, LHCb-PAPER-2022-012} from \pp and PbPb data, on cross-section measurements for exclusively produced \jpsi and $\Upsilonres(1S)$ in \PbPb collisions at $\sqsnn=5.44\tev$ and for exclusively produced \jpsi in \pPb\ collisions at $\sqsnn=8.8\tev$. 
As the published result on PbPb data is based on an integrated luminosity of $\lum = 0.2\invnb$, a dramatic improvement in precision will be achieved with Upgrade~II.

Additional measurements of interest in \PbPb collisions are the exclusive production of $\jpsi\gamma$ and $\Upsilonres\gamma$ pairs, $\eta_c$ mesons and $\chi_{c2}$ charmonium states. 
Their production proceeds through photon-photon interactions, and their study constraints the photon flux as well as the quarkonia production mechanism. 
Additionally, it would be interesting to search for exclusive $\chi_{c1}$ production, which could occur if one of the interacting photons is off-shell. All of these processes are characterised by a very small multiplicity in the detector, and are hence very efficiently reconstructed. 

During Upgrade~II data-taking, it is anticipated that the \lhc will be operating in parallel with the Electron-Ion Collider~(EIC) at Brookhaven National Laboratory in the USA.  A major focus of the EIC will be studies of the partonic structure of nuclei and potential signatures of gluon saturation~\cite{AbdulKhalek:2021gbh}. The complementarity of far-forward hadron collisions at \lhcb and deep inelastic scattering at the EIC will enable a new, precision understanding of nucleon structure.

\section{Probing the transition to deconfinement}
Collisions of large nuclei provide the most extreme conditions where relatively large volumes of quark-gluon plasma (QGP) can be studied in detail.  However, smaller collision systems allow the transition region between confinement and deconfinement to be explored, providing important constraints on theoretical models of the conditions necessary for plasma formation. 
Additionally, in collisions using highly aspherical nuclei such as neon or asymmetric collisions systems, the initial-state anisotropy due to the colliding nuclear shapes is imprinted on final-state observables and affects the hydrodynamic behaviour of the plasma. 
During LHC Run~3, short data-taking periods with collisions of oxygen and neon nuclei are being delivered to experiments, offering unique systems where such correlations can be investigated. 
These ongoing Run~3 studies will provide significant guidance on future directions, and are likely to motivate further data-taking with a range of different beam nuclei.
The \lhcb collaboration will take full advantage of the flexibility of beam species delivered by the \lhc in the Upgrade~II era, including:

\begin{itemize}
  \item Detailed studies of the emergence of potential QGP signatures in $pp$ collisions at HL-LHC energies;
  \item Measurements of the evolution of these signatures in $p$A collisions;
  \item Connections across $pp$, $p$A, and AA collisions as a function of system size, number of participating nucleons, and multiplicity.\end{itemize}

\section{Hydrodynamic properties of the plasma}

Measurements at RHIC and the LHC have proven that the QGP displays the properties of a strongly coupled fluid, which behaves like a perfect liquid with a shear viscosity near the quantum mechanical limit.  However, most of the measurements of the fluid properties of the plasma to date have focused on mid-rapidity, leaving the longitudinal expansion of the plasma relatively unexplored.  The far-forward acceptance of \lhcb allows this region to be probed in detail for the first time.  Existing \lhcb measurements of the plasma properties at forward rapidity show tension with model calculations that have been tuned to mid-rapidity measurements~\cite{LHCb-PAPER-2023-031}. Therefore, a full understanding of the QGP fluidity requires additional constraints that only \lhcb Upgrade~II can provide.

\begin{figure}%
    \centering
    \subfloat{{\includegraphics[width=0.49\textwidth]{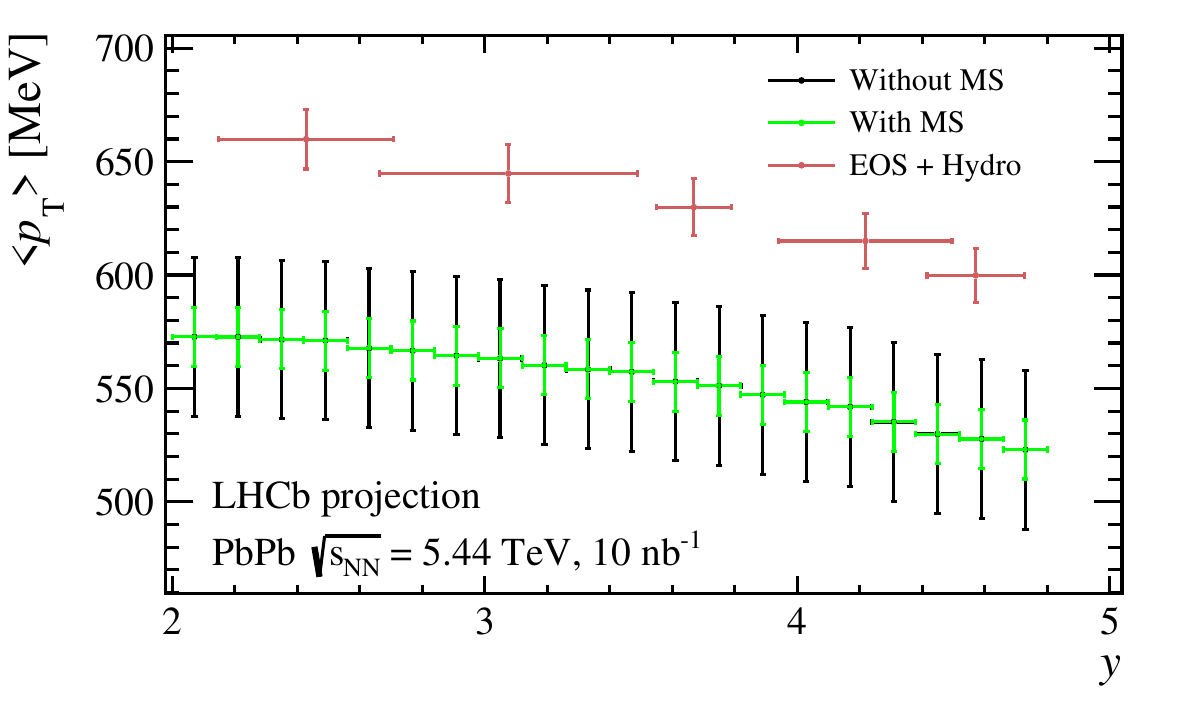} }}%
    \subfloat{{\includegraphics[width=0.49\textwidth]{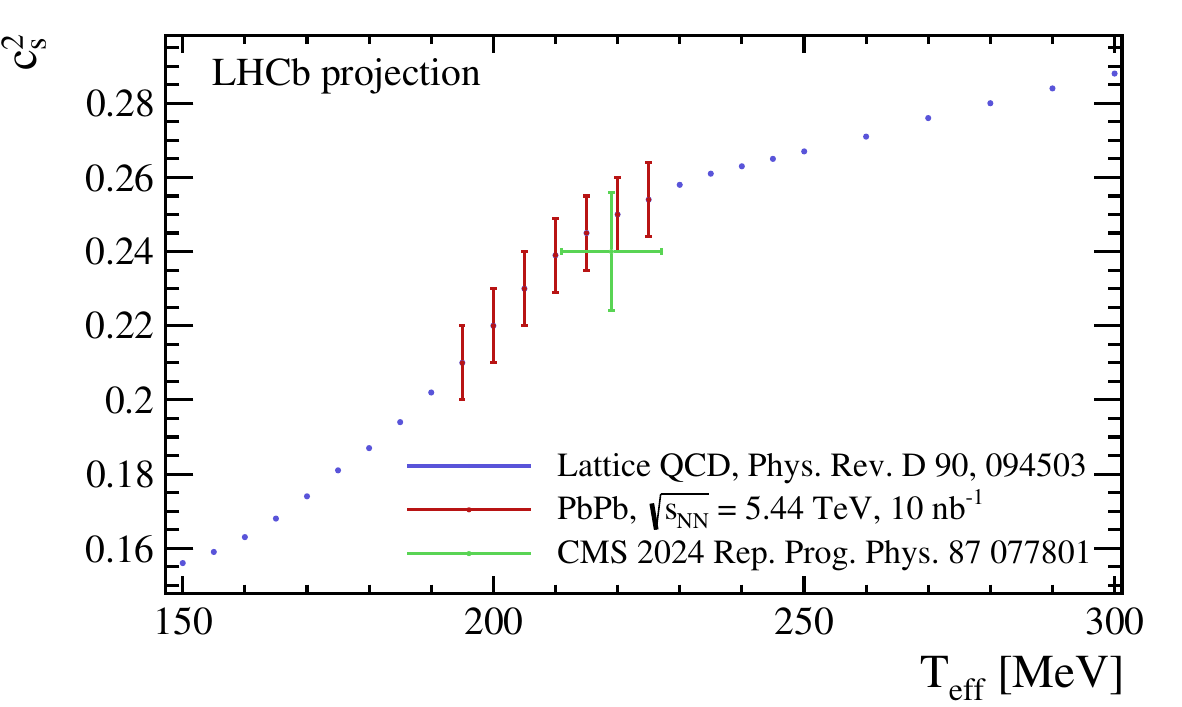} }}%
    \caption{Projections of (left) the mean transverse momentum $\langle\pt\rangle$ as a function of rapidity ($y$), and (right)~the plasma speed of sound $c_{s}^{2}$, with respect to the effective temperature of the plasma $T_{\rm eff}$, extracted from PbPb collisions with LHCb Upgrade~II.
    Modified from Ref.~\cite{Imanol_thesis}.}%
    \label{fig:bulk}%
\end{figure}

The upgraded \lhcb spectrometer will provide an ideal instrument for measuring the longitudinal hydrodynamic properties of the QGP at far forward rapidity.  In addition to detailed measurements of identified particle flow parameters, \lhcb will provide additional information on the fluid properties of the plasma.  Two examples are illustrated in Fig.~\ref{fig:bulk}. In the left plot, projections based on \pythia Angantyr simulation~\cite{Angantyr} for the measurement of unidentified particles average \pt as a function of rapidity are shown, and compared to a theoretical description~\cite{Nature_Jean_Yves} based on the QCD equation of state and on a hydrodynamic description of the plasma phase. 
A reduction of the experimental uncertainties by a factor of about 35\% is achieved through the addition of the new MS subdetector, which lowers the minimum accessible transverse momentum below 100\mev. 
The right plot shows the projected uncertainties for the plasma speed of sound as a function of the effective temperature, compared to lattice QCD calculations and a recent result by the CMS collaboration~\cite{CMS_speed_sound}. 
Thanks to the forward \lhcb acceptance, a precise measurement that covers a wide range in effective temperature will be accessible in \lhcb Upgrade~II.    

\section{Temperature of the plasma}
Production of charmonia and bottomonia, charm and bottom quark-antiquark bound states, has long been recognised as an outstanding probe to infer indirectly the temperature of the medium formed in heavy ion collisions~\cite{Matsui:1986dk}. Quarkonia are sensitive to multiple initial- and final-state effects in nuclear collisions, such as dissociation due to colour screening or interactions with comoving particles, which result in their suppression in large systems with respect to smaller ones. 
On the other hand, deconfined quarks in the plasma can recombine and form new quarkonia states at freeze out. Studying multiple quarkonia states as a function of the system size thereby provides insights into these effects. 

\begin{figure}[tb]
\centering
\includegraphics[width = 0.66\textwidth]{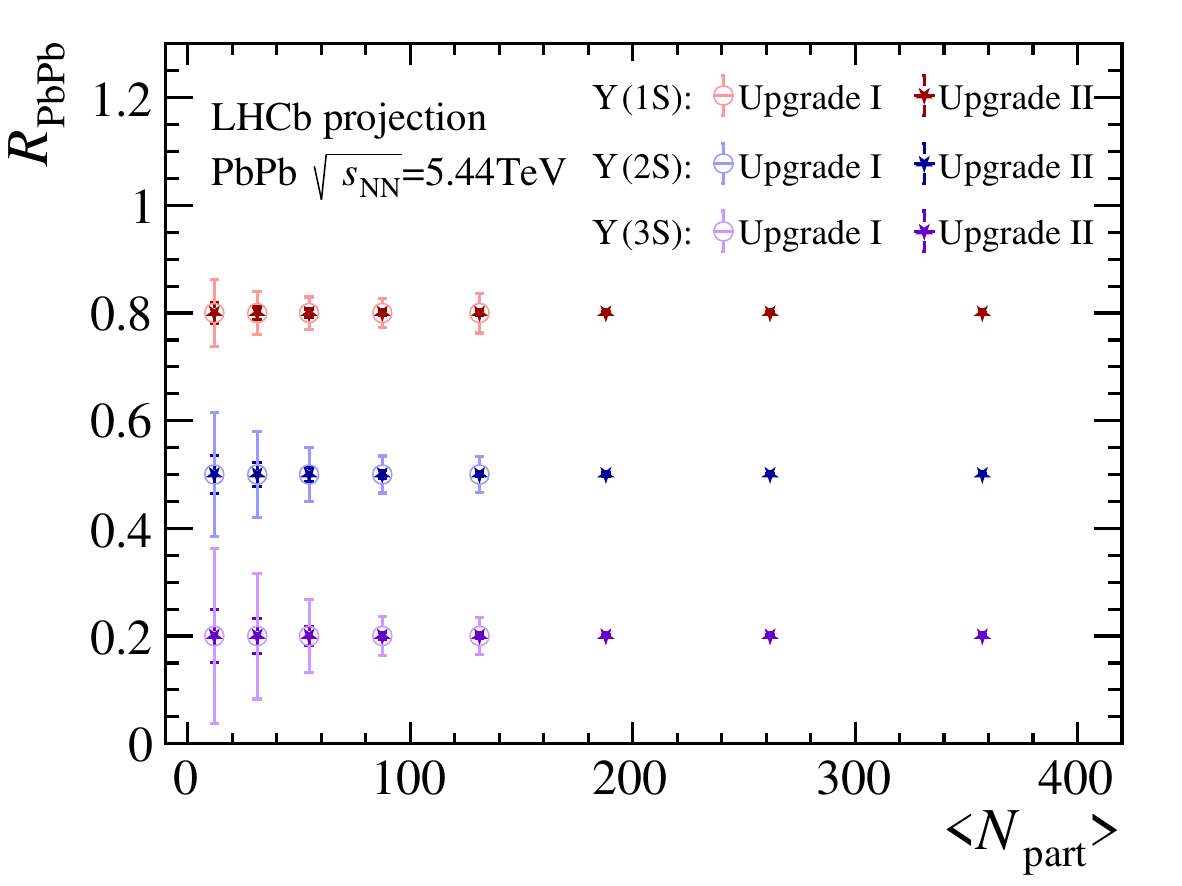}
\caption{Projected statistical uncertainties on the nuclear modification factors $R_{\PbPb}$ for \OneS, \TwoS and \ThreeS states in LHCb Upgrade~I with $\lum_{\mathrm{int}}=1\invnb$ and in LHCb Upgrade~II with $\lum_{\mathrm{int}}=10\invnb$ collected in \PbPb collisions at $\sqsnn=5.44\tev$.
The expected uncertainties are shown as functions of the number of participant nucleons $\langle N_{\rm part} \rangle$ for the $2<y<4.5$ and $\pt<20\gev$ kinematic range, and are estimated under the zero background hypothesis. 
To improve visibility, the absolute value of $R_{\PbPb}$ is shifted arbitrary to 0.8, 0.5 and 0.2 for the \OneS, \TwoS and \ThreeS, respectively, while the size of the error bars is maintained. }
\label{fig:quarkonia}
\end{figure}

A plethora of quarkonia production results are available from \lhcb in \pp, \pPb, peripheral \PbPb and fixed-target proton-nucleus (\textit{p}A) and lead-nucleus (PbA) collisions, mostly for $S$-wave vector states, such as the \jpsi, \psitwos, \OneS, \TwoS and \ThreeS~particles. The Upgrade~II detector will open new doors to extend the current studies. 
Leveraging the increased detector granularity, current measurements can be extended to central \PbPb collisions, where the produced medium has the largest extent and longest lifetime. 
Combined with the finer segmentation of the PicoCal (the upgraded electromgnetic calorimeter for LHCb Upgrade~II) compared to the current ECAL, and the additional acceptance for converted photons at low transverse momenta given by the MS tracker, feed-down contributions to the $S$-wave vector states, such as those from $P$-wave states like $\chi_{c/b}$ mesons, can be precisely accessed. 
This would overcome the limitation in the current interpretation of the quarkonium suppression, which might be only given in small systems by feed-down contributions. 
As an example, with the Run~1 and 2 detector \lhcb has measured the \decay{\chib}{\Upsilonres(nS)\gamma} feed-downs in \pp collisions~\cite{LHCb-PAPER-2014-031}, notably finding a $\approx40\%$ contribution from $\chi_b(3P)$ feed-down to the \ThreeS yield at $\pt^{\ThreeS}>25\gev$. 
As a similar \ThreeS suppression in high multiplicity \pp collisions has recently been reported by \lhcb~\cite{LHCb-PAPER-2024-038}, this could entirely be due to the $\chi_b(3P)$ contribution, which is expected to dissociate more easily given its low binding energy. 
Extending these measurements from \pp to more central \PbPb collisions and to lower \pt, which is only possible in the Upgrade~II conditions, is mandatory to draw a coherent picture for quarkonia production and suppression. 
Figure~\ref{fig:quarkonia} shows predictions for the expected uncertainties on measurements of the $\Upsilonres(nS)$ states nuclear modification factors as functions of the number of participant nucleons in Upgrade~I and Upgrade~II conditions, assuming samples corresponding to integrated luminosities of $\lum_{\mathrm{int}} = 1\invnb$ and $\lum_{\mathrm{int}} = 10\invnb$, respectively. 
A great improvement in precision with respect to the current detector is apparent, with increased reach in centrality.

Dilepton final states also provide powerful probes to constrain the temperature of the medium, as they play an important role in the understanding of the transition from an initial state of gluons to the hydrodynamics of a fluid in local equilibrium. The \lhcb detector has excellent capabilities for low-mass dilepton measurements. 
It has the best performance at the \lhc in \pp collisions for the rejection of the main background, the semileptonic decays of charm and beauty hadrons, as it combines very precise tracking and decay vertex reconstruction with excellent muon, electron, and hadron identification.  
Upgrade~II will provide the opportunity to extend this performance to central heavy-ion collisions~\cite{Turbide:2003si}.

\section{Transition back to confined matter (hadronization)}

As the plasma expands and cools, deconfined quarks that are near each other in phase space can coalesce to form hadrons at chemical freeze-out.  
As opposed to fragmentation in vacuum, this additional hadronization mechanism can result in enhanced production of baryons and exotic tetra- and pentaquark states in heavy ion collisions.  
Measurements of baryons and exotic hadrons containing heavy quarks are of particular interest. 
As the rate of heavy quark production is effectively set by the number of hard scatterings in the initial stages of the collision, variations of the fraction of heavy quarks that hadronize into mesons, baryons, and exotic hadrons give clear indications of changes to the hadronization process.

\begin{figure}[tb]
    \centering
    \subfloat{{\includegraphics[width=0.49\textwidth]{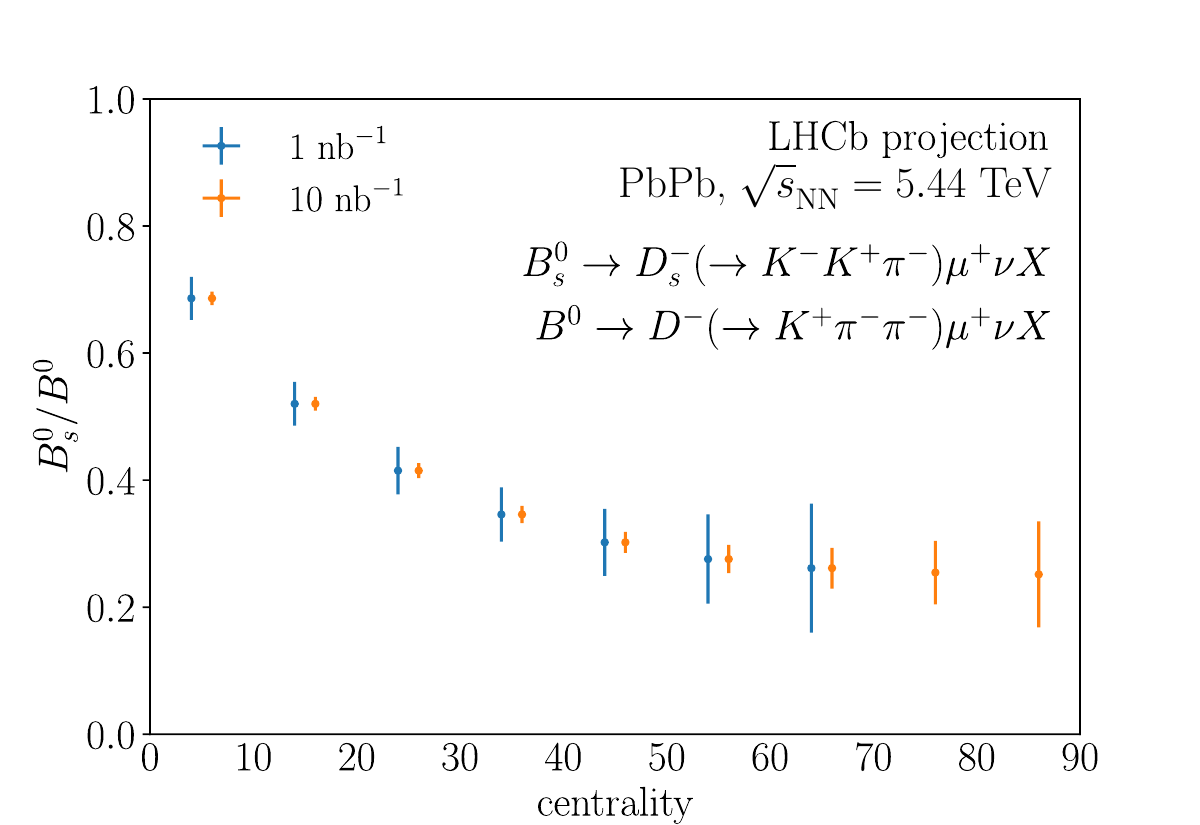} }}%
    \subfloat{{\includegraphics[width=0.49\textwidth]{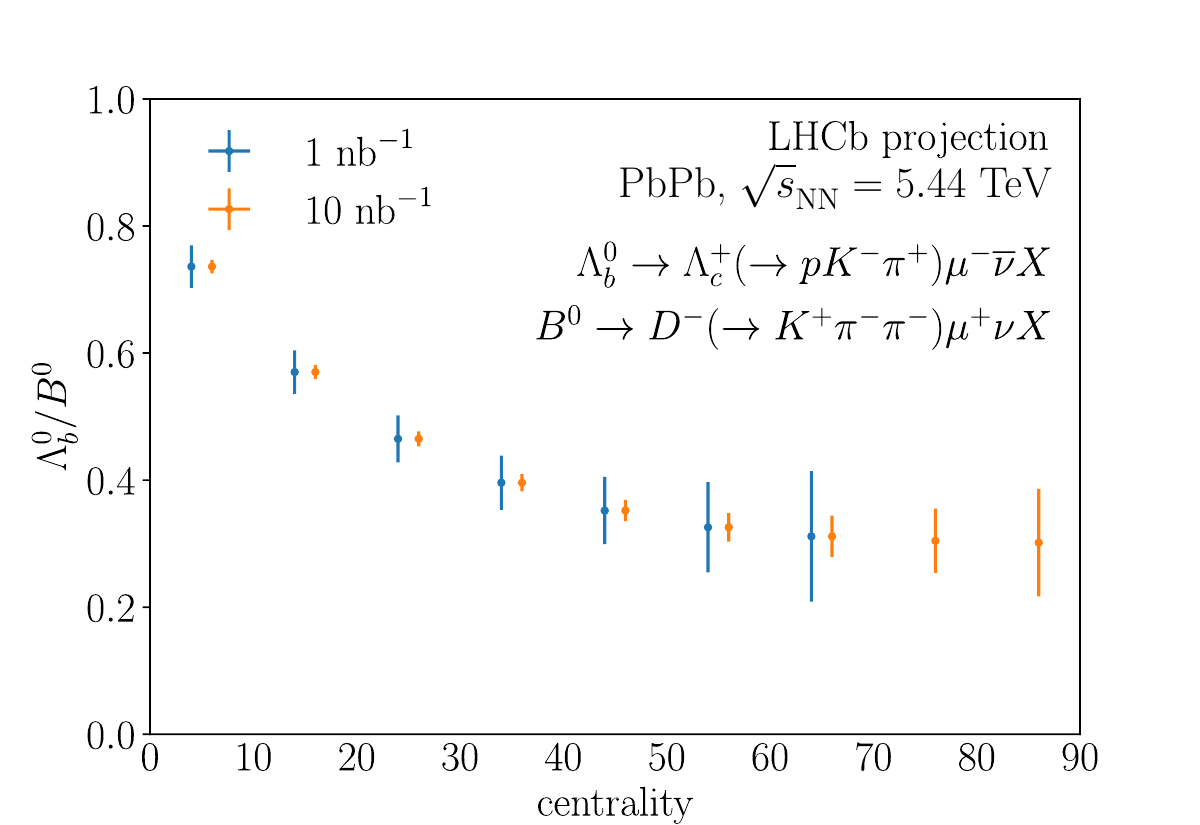} }}%
    \caption{Projected \lhcb Upgrade~II measurements of (left)~$\Bs/\Bz$ and (right)~$\Lb/\Bz$ ratios, as a function of centrality, for two different luminosity scenarios scaling from measurements in Ref.~\cite{LHCb-PAPER-2011-018}. 
    The projected uncertainties are statistical only.}%
    \label{fig:b_ratios}%
\end{figure}

Precise measurements of the production of \Bs mesons and \Lb baryons in \PbPb collisions are a priority of the LHCb Upgrade~II physics programme.  
The left and right panels of Fig.~\ref{fig:b_ratios} show projections for measurements of the ratio of \Bs mesons and \Lb baryons, respectively, to \Bz mesons based on scaling from measurements in Ref.~\cite{LHCb-PAPER-2011-018}. 
For the purposes of making these projections, the \Bs and \Lb enhancements are assumed to scale linearly with the number of binary nucleon-nucleon collisions.  
The large forward boost of $b$ hadrons at LHCb and the relatively high branching fraction of semi-leptonic decays allows for the most precise measurements of $b$ hadronization in heavy ion collisions.

Coalescence also provides a new mechanism for the formation of exotic multiquark states.  For many of these resonances, it is not immediately clear if they are composed of two conventional hadrons bound into a hadronic molecule, or whether they are compact objects composed of more than three valence quarks.  The exotic $\chi_{c1}(3872)$ state has been the focus of much recent attention, with various models of its production in PbPb collisions giving different conclusions depending on the nature of the state.  
Transport calculations favour the production of hadronic molecules over compact tetraquarks~\cite{Wu:2020zbx}, while calculations of quark coalescence expect the opposite~\cite{Yun:2022evm}.  
Statistical hadronization models cannot discriminate between compact and molecular scenarios, but expect a significant enhancement in the production of $\chi_{c1}(3872)$ and other exotic states~\cite{Andronic:2019wva}.  
Measurements of the $T_{cc}^{+}(3875)$ state are particularly interesting, as many models predict a large enhancement in central PbPb collisions with respect to \textit{pp}~\cite{Hu:2021gdg,Yun:2022evm,Abreu:2022lfy}.

\begin{figure}[tb]
\centering
\includegraphics[width = 0.9\textwidth]{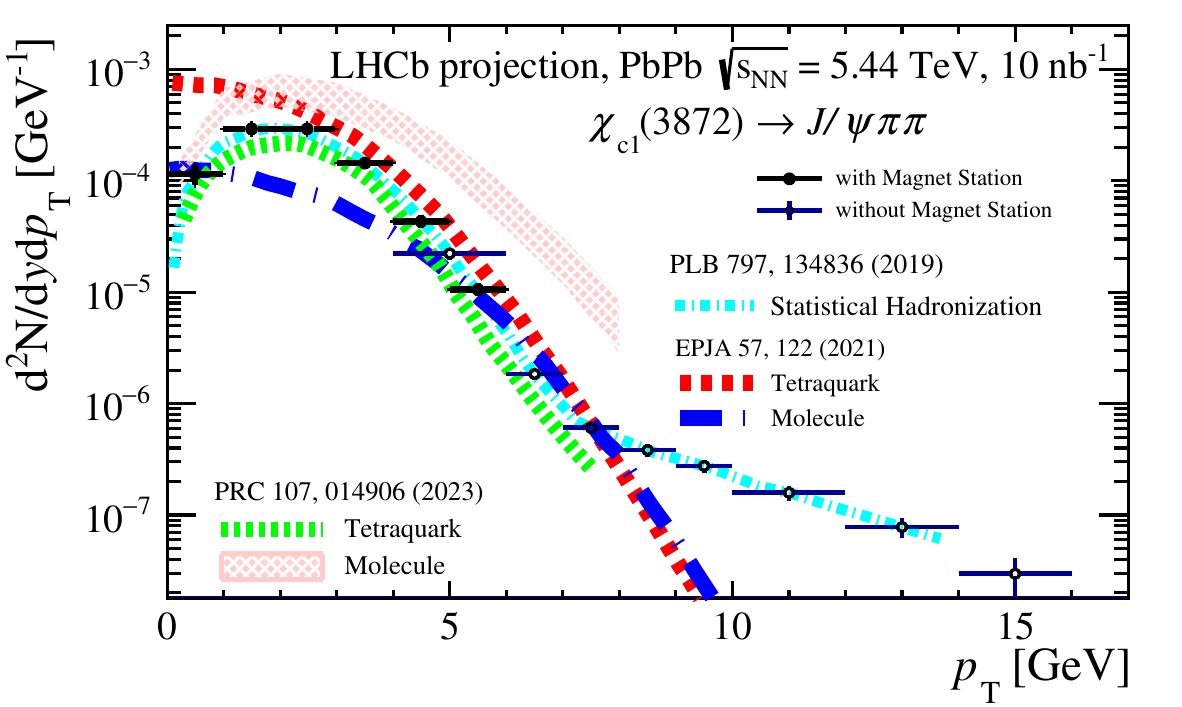}
\caption{Projected measurement of the yield of \mbox{$\decay{\theX}{\jpsi\pip\pim}$} decays in \PbPb collisions as a function of transverse momentum (\pt) over the rapidity interval $2<y<4.5$, assuming $10\invnb$ of collected luminosity with (filled circles) and without (open circles) the use of the Magnet Station tracker. Predictions from models of statistical hadronization~\cite{Andronic:2019wva}, coalescence~\cite{Yun:2022evm}, and transport~\cite{Wu:2020zbx} are shown for comparison.}
\label{fig:x3872_projections}
\end{figure}

These disagreements between theoretical predictions can be resolved by LHCb Upgrade~II.  
Figure~\ref{fig:x3872_projections} shows projections of the projected yield in $10 \invnb$ of PbPb collision data of the exotic $\chi_{c1}(3872)$ hadron, reconstructed in the $\jpsi\pip\pim$ decay channel, compared to various calculations.  
The forward boost at LHCb and the addition of the MS tracker allow the soft pions from  $\chi_{c1}(3872)$ decays to be reconstructed at very low $\pt$, where the models disagree significantly.  
These definitive measurements can only be accomplished with LHCb Upgrade~II.

\section{Conclusions}

A major focus of the LHCb collaboration in Run~5 and beyond is the exploration of nuclear matter at the highest density and temperature with $pp$, $p$A, and AA collisions at HL-LHC energies. With unique reach into the deep low-$x$ region of the nucleus, the most sensitive searches for gluon saturation can be performed at LHCb Upgrade~II.  The unprecedented range of mesons, baryons, and exotic hadrons that can be reconstructed by LHCb Upgrade~II will allow the QGP to be probed at different scales.  Measurements with LHCb Upgrade~II will have unique sensitivity to the nuclear initial state, the transition to deconfinement, the properties of the plasma, and  hadronization.  The instrumentation advances realized in the LHCb Upgrade~II detector have a direct impact on the LHC heavy ion program, enabling a unique suite of measurements that cannot be performed at any other experiment or facility.

\FloatBarrier
\addcontentsline{toc}{section}{References}
\bibliographystyle{LHCb}
\bibliography{main,standard,LHCb-PAPER,LHCb-CONF,LHCb-DP,LHCb-TDR}

\end{document}

%% file: preamble.tex
\usepackage[top=1in, bottom=1.25in, left=1in, right=1in]{geometry}

\usepackage{microtype}
\usepackage{lineno}  
\usepackage{xspace} 
\usepackage{caption} 

\usepackage{graphicx}  
\usepackage{color}
\usepackage{colortbl}
\graphicspath{{./figs/}} 

\usepackage{amsmath} 
\usepackage{amssymb}
\usepackage{amsfonts}
\usepackage{upgreek} 

\usepackage{multirow}

\newcommand*\patchAmsMathEnvironmentForLineno[1]{%
\expandafter\let\csname old#1\expandafter\endcsname\csname #1\endcsname
\expandafter\let\csname oldend#1\expandafter\endcsname\csname
end#1\endcsname
 \renewenvironment{#1}%
   {\linenomath\csname old#1\endcsname}%
   {\csname oldend#1\endcsname\endlinenomath}%
}
\newcommand*\patchBothAmsMathEnvironmentsForLineno[1]{%
  \patchAmsMathEnvironmentForLineno{#1}%
  \patchAmsMathEnvironmentForLineno{#1*}%
}
\AtBeginDocument{%
\patchBothAmsMathEnvironmentsForLineno{equation}%
\patchBothAmsMathEnvironmentsForLineno{align}%
\patchBothAmsMathEnvironmentsForLineno{flalign}%
\patchBothAmsMathEnvironmentsForLineno{alignat}%
\patchBothAmsMathEnvironmentsForLineno{gather}%
\patchBothAmsMathEnvironmentsForLineno{multline}%
\patchBothAmsMathEnvironmentsForLineno{eqnarray}%
}

\usepackage[pdftex,
            pdfauthor={\paperauthors},
            pdftitle={\paperasciititle},
            pdfkeywords={\paperkeywords},
]{hyperref}

\usepackage{hyperxmp}

\hypersetup{pdfcopyright={Copyright (C) \papercopyright}}
\hypersetup{pdflicenseurl={\paperlicenceurl}}

\usepackage[bottom,flushmargin,hang,multiple]{footmisc}

\usepackage[all]{hypcap} 

\input{lhcb-symbols-def} 

\usepackage{cite} 
\usepackage{mciteplus}

%% file: lhcb-symbols-def.tex
\usepackage{xspace} 
\usepackage{upgreek}

\def\lhcb   {\mbox{LHCb}\xspace}

\def\lhc    {\mbox{LHC}\xspace}

\def\MagUp {\mbox{\em Mag\kern -0.05em Up}\xspace}

\ifthenelse{\boolean{uprightparticles}}%
{

 \def\Ppi         {\ensuremath{\uppi}\xspace}

 \def\Pchi        {\ensuremath{\upchi}\xspace}                 
 \def\Ppsi        {\ensuremath{\uppsi}\xspace}

 \def\PDelta      {\ensuremath{\Delta}\xspace}                 
 \def\PXi         {\ensuremath{\Xi}\xspace}                 
 \def\PLambda     {\ensuremath{\Lambda}\xspace}                 
 \def\PSigma      {\ensuremath{\Sigma}\xspace}                 
 \def\POmega      {\ensuremath{\Omega}\xspace}                 
 \def\PUpsilon    {\ensuremath{\Upsilon}\xspace}
 \let\oldPi\Pi
 \def\PPi         {\ensuremath{\oldPi}\xspace}

 \def\PB      {\ensuremath{\mathrm{B}}\xspace}                 
                  
 \def\PD      {\ensuremath{\mathrm{D}}\xspace}

 \def\PJ      {\ensuremath{\mathrm{J}}\xspace}                 
 \def\PK      {\ensuremath{\mathrm{K}}\xspace}

 \def\Pb      {\ensuremath{\mathrm{b}}\xspace}

 \def\Pi      {\ensuremath{\mathrm{i}}\xspace}

 \def\Pp      {\ensuremath{\mathrm{p}}\xspace}

 \def\Ps      {\ensuremath{\mathrm{s}}\xspace}

 \def\thebaroffset{0.0em}
}
{

 \def\Ppi         {\ensuremath{\pi}\xspace}

 \def\Pchi        {\ensuremath{\chi}\xspace}                 
 \def\Ppsi        {\ensuremath{\psi}\xspace}                 
                  
 \mathchardef\PDelta="7101
 \mathchardef\PXi="7104
 \mathchardef\PLambda="7103
 \mathchardef\PSigma="7106
 \mathchardef\POmega="710A
 \mathchardef\PUpsilon="7107
 \mathchardef\PPi="7105
                  
 \def\PB      {\ensuremath{B}\xspace}                 
                  
 \def\PD      {\ensuremath{D}\xspace}

 \def\PJ      {\ensuremath{J}\xspace}                 
 \def\PK      {\ensuremath{K}\xspace}

 \def\Pb      {\ensuremath{b}\xspace}

 \def\Pi      {\ensuremath{i}\xspace}

 \def\Pp      {\ensuremath{p}\xspace}

 \def\Ps      {\ensuremath{s}\xspace}

 \def\thebaroffset{0.18em}
}
\newcommand{\offsetoverline}[2][\thebaroffset]{\kern #1\overline{\kern -#1 #2}}%

\makeatletter
\ifcase \@ptsize \relax
  \newcommand{\miniscule}{\@setfontsize\miniscule{4}{5}}
\or
  \newcommand{\miniscule}{\@setfontsize\miniscule{5}{6}}
\or
  \newcommand{\miniscule}{\@setfontsize\miniscule{5}{6}}
\fi
\makeatother

\DeclareRobustCommand{\optbar}[1]{\shortstack{{\miniscule (\rule[.5ex]{1.25em}{.18mm})}
  \\ [-.7ex] $#1$}}

\def\squark    {{\ensuremath{\Ps}}\xspace}

\def\bquark    {{\ensuremath{\Pb}}\xspace}

\def\pion   {{\ensuremath{\Ppi}}\xspace}

\def\pip    {{\ensuremath{\pion^+}}\xspace}
\def\pim    {{\ensuremath{\pion^-}}\xspace}

\def\KorKbar {\kern \thebaroffset\optbar{\kern -\thebaroffset \PK}{}\xspace}

\def\D       {{\ensuremath{\PD}}\xspace}

\def\DorDbar {\kern \thebaroffset\optbar{\kern -\thebaroffset \PD}\xspace}
\def\Dz      {{\ensuremath{\D^0}}\xspace}

\def\Dp      {{\ensuremath{\D^+}}\xspace}
\def\Dm      {{\ensuremath{\D^-}}\xspace}

\def\DpDm    {\ensuremath{\Dp {\kern -0.16em \Dm}}\xspace}

\def\B       {{\ensuremath{\PB}}\xspace}

\def\BorBbar {\kern \thebaroffset\optbar{\kern -\thebaroffset \PB}\xspace}
\def\Bz      {{\ensuremath{\B^0}}\xspace}

\def\Bd      {{\ensuremath{\B^0}}\xspace}

\def\BdorBdbar {\kern \thebaroffset\optbar{\kern -\thebaroffset \Bd}\xspace}

\def\Bs      {{\ensuremath{\B^0_\squark}}\xspace}

\def\BsorBsbar {\kern \thebaroffset\optbar{\kern -\thebaroffset \Bs}\xspace}

\def\jpsi     {{\ensuremath{{\PJ\mskip -3mu/\mskip -2mu\Ppsi}}}\xspace}
\def\psitwos  {{\ensuremath{\Ppsi{(2S)}}}\xspace}

\def\Upsilonres  {{\ensuremath{\PUpsilon}}\xspace}
\def\Y#1S{\ensuremath{\PUpsilon{(#1S)}}\xspace}
\def\OneS  {{\Y1S}\xspace}
\def\TwoS  {{\Y2S}\xspace}
\def\ThreeS{{\Y3S}\xspace}

\def\chib     {{\ensuremath{\Pchi_{b}}}\xspace}

\def\theX     {{\ensuremath{\Pchi_{c1}(3872)}}\xspace}

\def\proton      {{\ensuremath{\Pp}}\xspace}

\def\Lz          {{\ensuremath{\PLambda}}\xspace}

\def\LorLbar     {\kern \thebaroffset\optbar{\kern -\thebaroffset \PLambda}\xspace}

\def\Lb           {{\ensuremath{\Lz^0_\bquark}}\xspace}

\newcommand{\decay}[2]{\ensuremath{#1\!\to #2}\xspace} 

\def\to                 {\ensuremath{\rightarrow}\xspace}

\def\AT#1     {\ensuremath{A_{\mathrm{T}}^{#1}}\xspace}           

\def\C#1      {\ensuremath{\mathcal{C}_{#1}}\xspace}                       
\def\Cp#1     {\ensuremath{\mathcal{C}_{#1}^{'}}\xspace}                    
\def\Ceff#1   {\ensuremath{\mathcal{C}_{#1}^{\mathrm{(eff)}}}\xspace}        
\def\Cpeff#1  {\ensuremath{\mathcal{C}_{#1}^{'\mathrm{(eff)}}}\xspace}       
\def\Ope#1    {\ensuremath{\mathcal{O}_{#1}}\xspace}                       
\def\Opep#1   {\ensuremath{\mathcal{O}_{#1}^{'}}\xspace}                    

\newcommand{\aunit}[1]{\ensuremath{\text{\,#1}}}       

\newcommand{\tev}{\aunit{Te\kern -0.1em V}\xspace}
\newcommand{\gev}{\aunit{Ge\kern -0.1em V}\xspace}
\newcommand{\mev}{\aunit{Me\kern -0.1em V}\xspace}
\newcommand{\kev}{\aunit{ke\kern -0.1em V}\xspace}
\newcommand{\ev}{\aunit{e\kern -0.1em V}\xspace}
 
\newcommand{\mevc}{\ensuremath{\aunit{Me\kern -0.1em V\!/}c}\xspace}
\newcommand{\gevc}{\ensuremath{\aunit{Ge\kern -0.1em V\!/}c}\xspace}
\newcommand{\mevcc}{\ensuremath{\aunit{Me\kern -0.1em V\!/}c^2}\xspace}
\newcommand{\gevcc}{\ensuremath{\aunit{Ge\kern -0.1em V\!/}c^2}\xspace}

\def\cm   {\aunit{cm}\xspace}

\def\nb {\aunit{nb}\xspace}
\def\invnb {\ensuremath{\nb^{-1}}\xspace}

\def\sec  {\ensuremath{\aunit{s}}\xspace}

\def\gsim{{~\raise.15em\hbox{$>$}\kern-.85em
          \lower.35em\hbox{$\sim$}~}\xspace}
\def\lsim{{~\raise.15em\hbox{$<$}\kern-.85em
          \lower.35em\hbox{$\sim$}~}\xspace}

\def\sqsnn {\ensuremath{\protect\sqrt{s_{\scriptscriptstyle\text{NN}}}}\xspace}
\def\pt         {\ensuremath{p_{\mathrm{T}}}\xspace}

\newcommand{\lum} {\ensuremath{\mathcal{L}}\xspace}

\def\pythia     {\mbox{\textsc{Pythia}}\xspace}

\def\tell1  {TELL1\xspace}
\def\ukl1   {UKL1\xspace}

\newcommand{\eg}{\mbox{\itshape e.g.}\xspace}

\newcommand{\lhcborcid}[1]{\href{https://orcid.org/#1}{\hspace*{0.1em}\raisebox{-0.45ex}{\includegraphics[width=1em]{figs/orcidIcon.pdf}}}}


%% file: title-LHCb.tex

\begin{titlepage}
\pagenumbering{roman}

\vspace*{-1.5cm}
\centerline{\large EUROPEAN ORGANIZATION FOR NUCLEAR RESEARCH (CERN)}
\vspace*{1.5cm}
\noindent
\begin{tabular*}{\linewidth}{lc@{\extracolsep{\fill}}r@{\extracolsep{0pt}}}
\ifthenelse{\boolean{pdflatex}}
{\vspace*{-1.5cm}\mbox{\!\!\!\includegraphics[width=.14\textwidth]{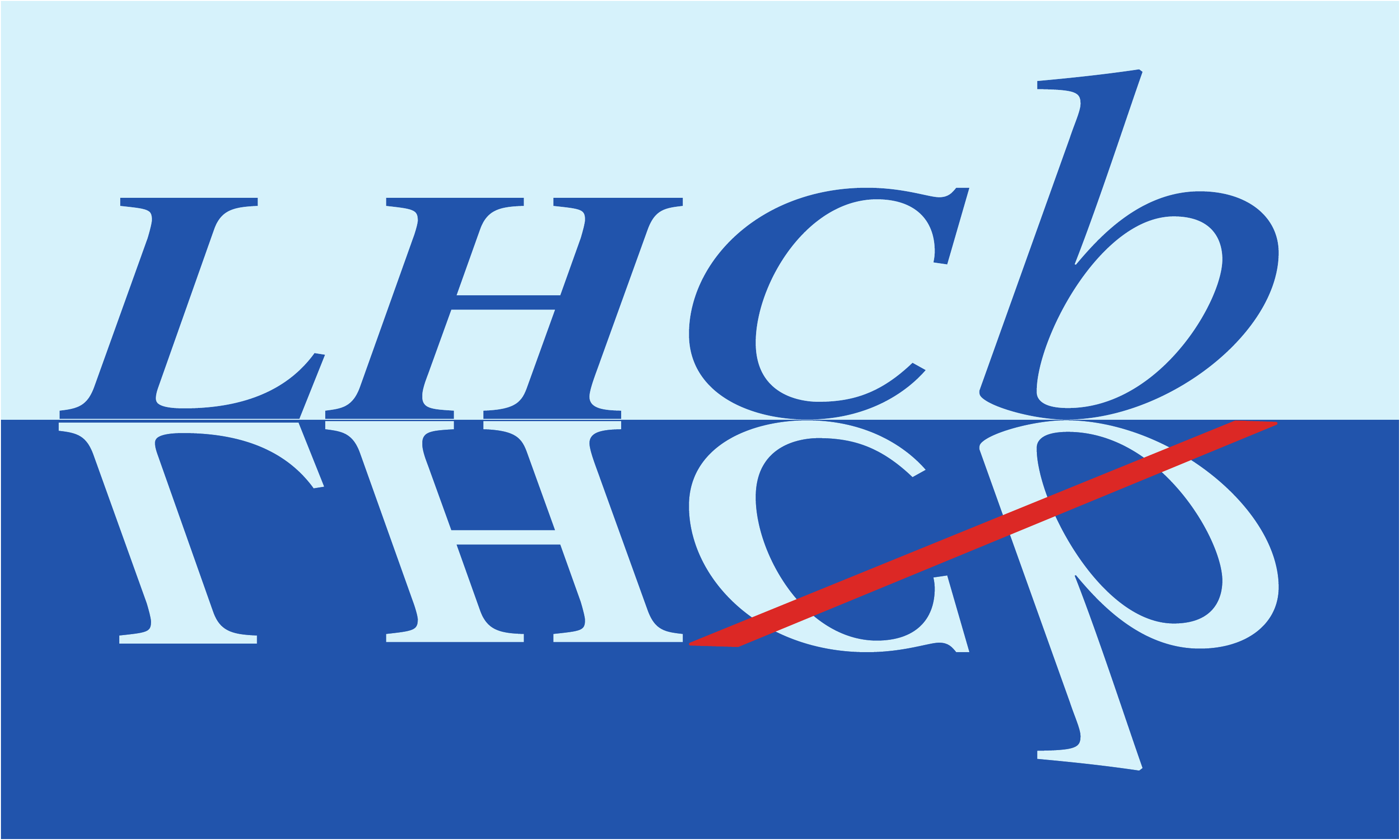}} & &}%
{\vspace*{-1.2cm}\mbox{\!\!\!\includegraphics[width=.12\textwidth]{figs/lhcb-logo.eps}} & &}%
\\
 & & LHCb-PUB-2025-003 \\  
 & & \today \\ 
 & & \\
\end{tabular*}

\vspace*{2.0cm}

{\normalfont\bfseries\boldmath\huge 
\begin{center}
    \papertitle \\
    \vspace*{0.5cm}
  {\normalsize Input to the European Particle Physics Strategy Update 2024--26}
\end{center}
}

\vspace*{1.0cm}

\begin{center}
\paperauthors\footnote{
    Contact authors: 
    Vincenzo Vagnoni (\href{mailto:vincenzo.vagnoni@cern.ch}{vincenzo.vagnoni@cern.ch}),
    Tim Gershon (\href{mailto:tim.gershon@cern.ch}{tim.gershon@cern.ch}),
    Giovanni Punzi (\href{mailto:giovanni.punzi@cern.ch}{giovanni.punzi@cern.ch})
}
\end{center}

\vspace{\fill}

\begin{abstract}
    \noindent
    A second major LHCb detector upgrade will be installed during long shutdown 4 (LS4) of the CERN Large Hadron Collider.
    The new detector will provide excellent performance for studies of Quantum Chromodynamics at high temperature and density, as achieved in collisions of heavy nuclei.  The high granularity of the tracking system will allow lead-lead collisions to be reconstructed across the full range of centrality at far forward rapidity for the first time.  Moreover, the forward acceptance of the detector, covering the pseudorapidity region close to the beamline, and the capability to reconstruct a wide range of hadrons containing strange, charm, and beauty quarks result in unique potential to probe the medium produced in the collisions.  In this document, the heavy ion physics programme that will be pursued in LHCb Upgrade~II is summarised, including precision studies of the partonic structure of nuclei, probes of the conditions allowing deconfinement, and measurements of plasma properties. 
\end{abstract}

\vspace{\fill}

{\footnotesize 
\centerline{\copyright~\papercopyright. \href{\paperlicenceurl}{\paperlicence}.}}
\vspace*{2mm}

\end{titlepage}


\newpage
\setcounter{page}{2}
\mbox{~}